\documentclass[twocolumn,showpacs,preprintnumbers,amsmath,amssymb]{revtex4}

\usepackage[unicode]{hyperref}
\usepackage[caption=false]{subfig}
\usepackage{graphicx}
\graphicspath{ {./images/} }
\usepackage{dcolumn}
\usepackage{bm}

\begin{document}

\title{Recent development on fragmentation, aggregation and percolation
}%

\author{M. K. Hassan 
}%
\date{\today}%

\affiliation{
University of Dhaka, Department of Physics, Theoretical Physics Group, Dhaka 1000, Bangladesh \\
}

\begin{abstract}
In this article,  I have outlined how an accomplished researcher like Robert Ziff
has influenced a new generation of researchers across the globe like gravity as an action-at-a-distance. 
In the 80s Ziff made significant contributions to the kinetics of fragmentation 
followed by the kinetics of aggregation. Here, I will discuss fractal and multifractal that emerges
in fragmentation and aggregation processes where the dynamics is governed by non-trivial conservation laws.
I have then discussed my recent works and results
on percolation where I made extensive use of Newman-Ziff fast Monte Carlo algorithm. To this end, I have defined entropy which paved the way to define specific heat and show that the critical
exponents of percolation obey Rushbrooke inequality. Besides, we discuss how entropy
and order parameter together can help us to check whether the percolation is
accompanied by order-disorder transition or not. The idea of entropy also help to explain
why encouraging smaller cluster to grow faster than larger clusters makes the transition explosive.

\end{abstract}


\maketitle

\section{introduction}

Since the 80s Robert Ziff has been active in research and teaching
while at the same time accomplishing path-breaking work on a wide range of problems 
in statistical mechanics. I have been so inspired by his work that my research 
track is like following the footsteps of his works. I started my research career with kinetics of fragmentation in which Ziff in the 80s made significant contributions in finding exact solutions under
various different conditions \cite{ref.ziff_frag_1,ref.ziff_frag_2,ref.ziff_frag_3}. 
In fact, my PhD thesis was on the kinetics of fragmentation where I extended the problem to describe fragmentation of particles of higher dimensions \cite{ref.hassan_rodgers}.
During the same time I also worked on the random sequential adsorption process \cite{ref.hassan,ref.hassan_rsa1, ref.hassan_rsa2}. In between, I 
worked on kinetics of aggregation and shown for the first time that like in fragmentation, aggregation
too can result in the emergence of fractal under certain conditions \cite{ref.fractal_cda_1,ref.fractal_cda_2,ref.fractal_self}. For instance, we describe
fragmentation processes by fractal if the mass of the system decreases with time following 
a power-law while
in aggregation processes it is only true if the mass increases in the same fashion. In other
words, fractal in kinetics of fragmentation and aggregation is only possible if the conservation
of total mass is violated.

However the most significant and extensive part of his research has been on the theory of percolation. 
Newman-Ziff algorithm has inspired a generation of researchers like myself to study percolation
\cite{ref.per_ziff_0,ref.per_ziff_algorithm}.
Percolation is perhaps one of the simplest models to study phase transition and critical phenomena. 
Yet, an exact and rigorous analytical solution is only possible in one dimension and in infinite 
dimensions like on Bethe lattice. However, Ziff has contributed significantly in finding 
some analytical solutions to percolation, both on the lattice level and in the continuum limit, though
not based on first principle which still is an open problem  \cite{ref.per_ziff_perco_1,ref.per_ziff_perco_2,ref.per_ziff_perco_3,ref.per_ziff_perco_4,
ref.per_ziff_perco_5}. Moreover, despite having
been a simple model for phase transition for more than 60 years, yet we did not know how to 
define entropy for percolation until 1999 albeit their entropy is not consistent with order
parameter \cite{ref.tsang, ref.tsang_1,ref.Vieira}. Entropy, along with order parameter, are two of the key quantities 
which are used to define the order of transition. Once we know how to define entropy, 
we can immediately find specific heat which has long been an elusive quantity too.

To define percolation we need two things (i) a skeleton and (ii) the rules of the game. Since
1998 with the invention of scale-free and small-world networks, the use of network as a skeleton
in percolation and in other physics models gained a lot of attention. 
To this end, Achlioptas {\it et al.} in 2009 proposed a variant of percolation  \cite{ref.Achlioptas, ref.raissa}. 
The question they asked was: What if
we pick two candidate links instead of one and ultimately add only the one that minimizes
the resulting cluster size? Using Erd\"{o}s and R\'{e}nyi's random network as a skeleton 
\cite{ref.erdos}, Achlioptas {\it et al.} showed that such a small change in the rules of percolation makes
a huge impact in the way giant cluster appears in the system. Indeed, the order 
parameter $P$ undergoes such an abrupt 
transition at the critical point $t_c$, where the relative link density $t$ is the ratio
of the number of links $n$ added and the network size $N$, 
that it was at first thought to be a discontinuous transition. 
In support of their claim, Achlioptas {\it et al.} measured the difference in number of links added 
$\Delta=n_2-n_1$ between the last step $n_1$ for which the largest cluster $S_{{\rm max}} <N^{1/2}$ and 
the first step $n_2$ for which $S_{{\rm max}}>N/2$. They showed that it is not an 
extensive quantity rather they found $\Delta\sim N^\kappa$ with $\Delta/N\rightarrow 0$ 
in the thermodynamic limit which is a clear sign of first order transition. 
The claim that explosive percolation (EP) on ER describes first order transition 
resulted in a series of articles, some 
supporting the claim and some against it \cite{ref.Friedman, ref.ziff_1, ref.radicchi_1, ref.Costa_2, ref.souza, ref.cho_1,  ref.ara, ref.nagler}.
However, finally in 2011 it was agreed that 
EP transition is actually continuous but with first order like finite-size effects.\cite{ref.da_Costa, ref.Grassberger,ref.Riordan, ref.Tian}.

The most cited work of Robert Ziff is on the kinetic phase transition in an irreversible 
surface-reaction model which is now well-known as the Ziff-Gulari-Barshad model \cite{ref.ZGB}. 
Besides this, the Newman-Ziff algorithm 
in percolation theory, the shattering transition in fragmentation and the gelation transition in aggregation,  have been his path breaking works \cite{ref.per_ziff_0,ref.per_ziff_algorithm,ref.shattering,ref.gelation}. The rest of the article is organized as follows.  
I focus on his works in a chronological order and alongside 
I discuss about my contributions in those respective areas. However, the main focus
will remain on percolation theory as it is my current field of research and I am working extensively 
on it. To this end, in sec I we discuss kinetics of fragmentation of one dimensional and higher dimensional particles. I have found some new and exciting results. Besides, I am working on giving a thermodynamic formalism
of percolation theory.

\section{Kinetics of fragmentation}


Kinetics of fragmentation is an important physical phenomena that occurs in numerous physical, chemical
and geological processes. Examples include droplet breakup, fibre length reduction, polymer
degradation, rock crushing and grinding \cite{ref.droplet, ref.fiber_reduction, ref.polym_degrad_1,ref.polym_degrad_2}. Kinetics of fragmentation is a stochastic process 
that describes how the particle size distribution function $c(x,t)$, where $c(x,t)dx$ is 
the number of particles in the size range $x$ and $x+dx$ at time $t$, evolves with time
according to the following equation
\begin{eqnarray} 
\label{eq:binery}
{{\partial c(x,t)}\over{\partial t}}&=&-c(x,t)\int_0^x dyF(x-y,y)\nonumber\\
&+&2\int_x^\infty dyc(y,t)F(x,y-x).
\end{eqnarray}
Here, $F(x,y)$ is called the fragmentation kernel that captures the details of how a parent 
particle of size $(x+y)$ breaks into two daughter particles of
sizes $x$ and $y$ \cite{ref.ziff_frag_3}. The first term on the right-hand side
of Eq.~(\ref{eq:binery}) describes the loss of particles of size-$x$  due to their splitting into smaller ones, while the second term describes the gain of particles of size-$x$ from the fragmentation of 
$y>x$ particles. Ziff and McGrady have obtained exact solution of this master equation for many 
different choices of the fragmentation kernel both for its discrete and continuum version
\cite{ref.ziff_frag_1,ref.ziff_frag_2, ref.ziff_frag_3}. To this 
end, they used some unique and innovative methods to solve the equation which I found extremely 
useful over my entire research career. In 1987  Ziff and McGrady reported that the fragmentation equation exhibits shattering transition \cite{ref.shattering}. In this transition the smaller particles break up at 
increasingly rapid rates, resulting in mass being lost to a phase of ‘‘zero’’-size particles. In
some sense it is reminiscent of the Bose Einstein condensation. I kept reading and reading
all these inspiring articles and at the same time I was thinking of new ideas. Within a few months of 
working on these papers we, me and my supervisor, published our first paper 
\cite{ref.hassan_multi_frag_4}. 

In our first paper on fragmentation, we extended the fragmentation equation to higher dimensions. 
We thought we were the first to work along that direction at that time. Later we found out that there
were in fact two other groups (Tarjus et al and Krapivsky et al.) who were also working on the
same problem independently \cite{ref.krapivsky,ref.boyer_tarjus}. Surprisingly, three 
groups independently extended the fragmentation equation into higher dimensions and 
all the three articles submitted within March and July 1994 which were published in the
Physical Review E \cite{ref.hassan_rodgers, ref.krapivsky,ref.boyer_tarjus}. 
Boyer, Targus and Viot addressed the shattering aspect of the problem, Krapivsky
and Ben-Naim addressed scaling and multiscaling and we focused on exact solutions. 
One of the surprising results that Krapivsky and Ben-Naim reported was that in higher dimensions,
apart from the trivial conservation of mass principle, the system is governed by infinitely many non-trivial conservation laws.  

In 1994, Krapivsky and Ben-Naim used the spirit of Cantor set in the master 
equation for fragmentation process and shown that the resulting equation violates the usual mass 
conservation which is however replaced by another non-trivial conservation law namely the 
$d_f$th moment of $c(x,t)$ \cite{ref.krapivsky_naim_fractal}. Later, we extended the idea to 
the dyadic Cantor set and to the random sequential deposition of a mixture of particles whose size
distribution follows a power-law. To describe dyadic Cantor set we just have to replace
the factor $"2"$ in the gain term by $1+p$ so that at each time step, one of the two newly created 
fragments is removed from the system with probability $1-p$. 

In 1986  Ziff and McGrady proposed a model where particles are more likely to break in the center 
than on either end \cite{ref.ziff_frag_1}.  In 1995, we generalized it by choosing Gaussian rate kernel 
\begin{equation}
   F(x-y,y)=\frac{(x-y)^{\beta}y^{\beta}}{B(\beta+1,\beta+1)}
\end{equation}
where
we used a parameter $\beta$ that could control the degree of randomness such that for $\beta>0$ 
particles are more likely 
to break in the middle than on either end and in the limit $\beta\rightarrow \infty$ particles 
are only broken in the middle \cite{ref.fractals_95,ref.fractals_2002}. We have shown that fractal dimension 
increases with increasing $\beta$.

\begin{figure}

\centering

\subfloat[]
{
\includegraphics[height=7.0 cm, width=7.5 cm, clip=true]
{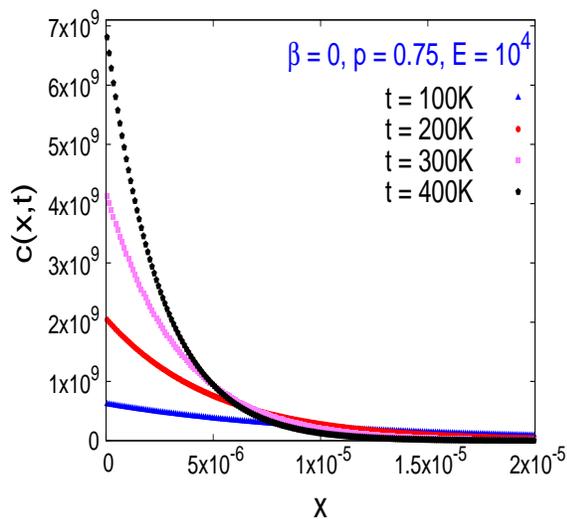}
\label{fig:1a}
}

\subfloat[]
{
\includegraphics[height=7.0 cm, width=7.5 cm, clip=true]
{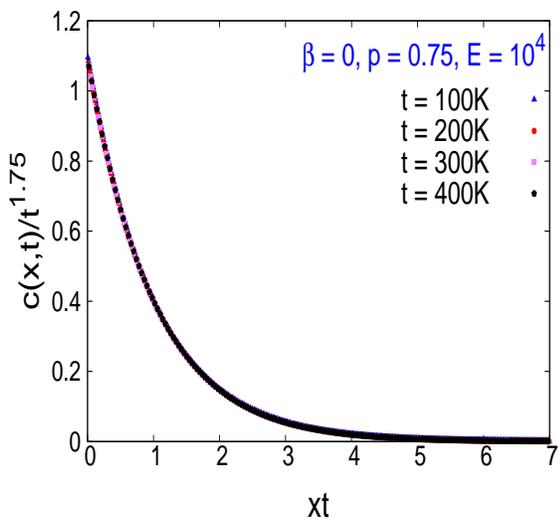}
\label{fig:1b}
}

\caption{(a) The distribution function $c(t,x)$ is drawn as a function of $x$  
for different time ($t=100,200,300,400$) with fixed $\beta=0$ and $p=0.75$ where $p$ is the probability with which the 
newly created fragment is kept in the system.
(b) The same set of data set  are used to plot them in the self-similar scale and find excellent data collapse. Here, all the plots represent ensemble average over $E=1\times 10^4$ independent realizations.  
} 
\label{fig:1ab}
\end{figure}

\begin{figure}

\centering

\subfloat[]
{
\includegraphics[height=7.0 cm, width=7.5 cm, clip=true]
{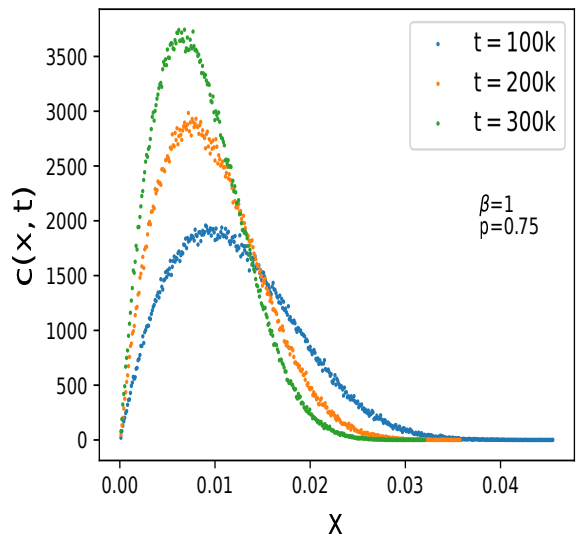}
\label{fig:2a}
}

\subfloat[]
{
\includegraphics[height=7.0 cm, width=7.5 cm, clip=true]
{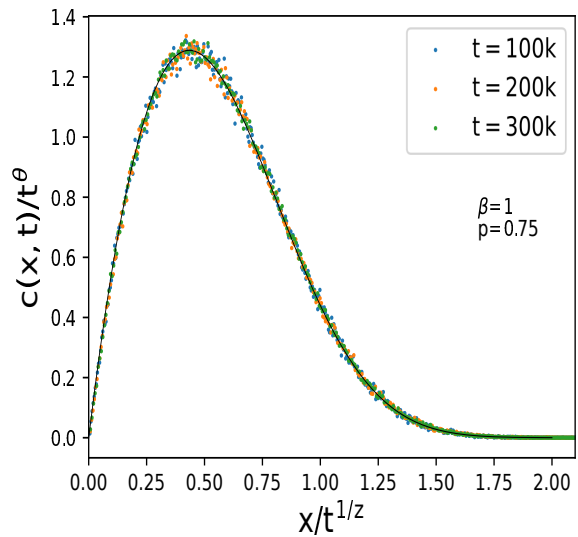}
\label{fig:2b}
}

\caption{(a) The distribution function $c(t,x)$ is drawn as a function of $x$ for 
$\beta=1$. Each plot represents data obtained from snapshots taken at different time
$t=100,200,300$. (b) The same set of data are used to plot them in the self-similar 
scale and find excellent data collapse revealing that $c(x,t)$ exhibits dynamic scaling. 
} 

\label{fig:2ab}
\end{figure}

I continued working on one dimensional fragmentation equation primarily the case where
mass conservation is violated. We found that the violation of mass conservation is always
accompanied by the emergence of stochastic fractal. On the hand, one of the essential
criterion of fractal is self-similarity. In the case of stochastic fractal, self-similarity
means that the distribution function $c(x,t)$ exhibits dynamic scaling
\begin{equation}
\label{eq:dynamic_s_fragmentation}
    c(x,t)\sim t^{\theta z}\phi(x/t^{-z}),
\end{equation}
where $\phi(\xi)$ is the scaling function, $\theta$ is the mass exponent and $z$ 
is the kinetic exponent \cite{ref.family_Vicsek,ref.scaling}. Note that the mass 
exponent $\theta=1+d_f$ where $d_f$ is the fractal dimension. It means that
 the plots of $c(x,t)$ versus $x$ plotted from the snapshots of the system taken at different
 time will be distinct. However, all these distinct plots collapse into a single universal
 scaling function if we plot $c(x,t)t^{-\theta z}$ versus $x/t^{-z}$ instead. In 
 Figs. (\ref{fig:1a}) and (\ref{fig:2a}) we first plot distribution function  $c(x,t)$ as a function of
$x$ for dyadic Cantor set for $\beta=0$ and $\beta=1$ respectively to see how it varies with $\beta$ value \cite{ref.fractal_pandit,ref.fractal_rakib}. In Figs. (\ref{fig:1b}) and (\ref{fig:2b}) we use the same data to
plot $c(x,t)t^{(1+d_f)z}$ versus $x/t^{-z}$ and find excellent data collapse
of all the distinct
plots in Figs. (\ref{fig:1a}) and (\ref{fig:2a}) respectively. They suggest that the snapshots
taken at different times are similar, which is also a kind of symmetry in continuous time. 
It has been shown that the fractal dimension $d_f$ increases with $\beta$ and for 
all $\beta$, the $d_f$th moment is always a conserved quantity \cite{ref.fractals_2002}. 
We have recently shown a connection between
these conserved quantities and Noether's theorem that states that where a continuous symmetry exists
there must exist a conserved quantity \cite{ref.fractal_rakib}. 

Fragmentation equation never stopped giving surprising results. In 1996, we show that in 
planar fragmentation each non-trivial conserved quantity can be used as a multifractal 
measure \cite{ref.multifractality_pla, ref.multifractality_pre}.  
 Besides this, the planar fragmentation can also be described as random sequential division of plane into mutually
exclusive rectangular cells which we regarded as the weighted planar stochastic lattice (WPSL)
\cite{ref.hassan_njp,ref.hassan_jcp}. 
One of the most interesting findings of this work is the emergence of infinitely many conserved
quantities $\sum_i x_i^{m-1}y_i^{4/m-1}=1$ $\forall \ m$. Except $m=2$ case or the total 
mass conservation, all the other conserved quantities are highly non-trivial
whose existence we only know because we can solve the problem analytically. A far more
interesting fact is that each of the non-trivial conserved quantities can be regarded
as the multifractal measure such the $i$ cell contains only $\mu_i(m)=x_i^{m-1}y_i^{4/m-1}$ of the total measure.
The distribution of this measure can be best described
as a multifractal. Since each of the infinitely many non-trivial conserved quantities 
can be a measure, there are thus infinitely many multifractal $f(\alpha)$ spectra
where $\alpha$ is known as the H\"{o}lder exponent. Interestingly,
if we replace the center of each cell of the WPSL by a node and common border between cells by a link
between the corresponding node then it emerges as a scale-free network. More recently, we have 
solved a class of models where by dividing the plane vertically or horizontally with equal
probability the resulting network is not only scale-free with smaller exponent of the power-law degree distribution but also small-world \cite{ref.tushar_hassan_1,ref.tushar_hassan_2}. It is small-world because we find that the mean geodesic path length increases logarithmically with
system size and the total mean clustering coefficient is high and independent of system size. It implies that
it is also a small-world network. 

\section{Kinetics of aggregation}

Yet another field of research where Ziff studied extensively and contributed enormously is the kinetics
of aggregation or polymerization \cite{ref.ziff, ref.ziffEtAl}. The most successful equation that can describe the kinetic of 
aggregation process is the Smoluchowski equation
\begin{eqnarray}
\label{eq:smoluchowski}
& & {{\partial c(x,t)}\over{\partial t}}= -c(x,t)\int_0^\infty K(x,y)c(y,t)dy 
\nonumber \\ & + &{{1}\over{2}} \int_0^x K(y,x-y) c(y,t)c(x-y,t)dy,
\end{eqnarray} 
where $K(x,y)$ is the aggregation kernel that describes the rate at which particle
$x$ and $y$ meets \cite{ref.smoluchowski}.
The first (second)
term on the right hand side of Eq. (\ref{eq:1}) describes the loss (gain) of size $x$ 
due to merging of size $x$ ($(x-y)$) with particle of size $y$.Unlike fragmentation equation, 
finding exact 
solution for various different choices of aggregation kernel $K(x,y)$ is still a formidable task. Robert Ziff focused mostly on long time limit or scaling
solution. One of the most striking results is that the concentration $c(x,t)$ of particles of size $x$ at time $t$ exhibits dynamic scaling
\cite{ref.gelation,ref.scaling}. Like shattering transition in fragmentation, Ziff has shown that 
for certain choice of aggregation kernel the Smoluchowski
equation describes gelation transition \cite{ref.gelation,ref.ziff_stell,ref.ziff_ernst}. It is a phase where the system loses its mass to gel phase.
The sol-gel transition is a very interesting field of study due to its own right. His work led
to extensive study of the Smoluchowski equation. 

While I was working on my Phd on the kinetics of fragmentation, I was also reading articles on aggregation 
especially the articles authored by Ziff, Redner and Ernst. While reading articles and books, 
I realized that most numerical and experimental works suggest that almost 
always aggregates can be best described as fractal \cite{ref.vicsek,ref.lin,ref.weitz}. 
However, despite extensive studies, there did not exist any analytically solvable model within
the framework of Smoluchowski equation which can support this geometric aspect. Finding
a model that can describe the emergence of fractal could help us 
know why the fractal is ubiquitous in the aggregation process. In 2008-2013, I was finally
successful to find two different aggregation models within the framework of
Smoluchowski equation which can account for the emergence of fractal
\cite{ref.fractal_cda_1,ref.fractal_cda_2,ref.fractal_self}. Interestingly, it has been found that fractal in fragmentation
is only possible if the mass of the system is either removed or the size of 
the particles are continuously
decreased, for instance by evaporation, while fractal in aggregation emerges only if mass is added or particles are continuously grown, say, by condensation.
In general, for the emergence of fractal we must have a system which is open. 

The first model was on aggregation of continuously growing particles by 
heterogeneous condensation which we solved exactly in one dimension \cite{ref.fractal_cda_1}.  
 The generalized Smoluchowski (GS) equation that can describe condensation-driven aggregation is given by
\begin{eqnarray}
\label{eq:1}
\Big[{{\partial }\over{\partial t}} &+&  {{\partial}\over{\partial x}} v(x,t) \Big]c(x,t)
=-c(x,t)\int_0^\infty K(x,y)c(y,t)dy \nonumber \\ &+ & {{1}\over{2}}\int_0^x dy K(y,x-y)
c(y,t)c(x-y,t),
\end{eqnarray}
The second term on the left hand side of the above equation accounts for the growth by 
condensation with velocity $v(x,t)$. 
However, the GS equation can only describe the condensation-driven aggregation (CDA) 
model if the growth velocity $v(x,t)$, 
the collision time $\tau$, and the kernel $K(x,y)$ are suitably chosen. In the absence of the 
second term on the right hand side,
Eq. (\ref{eq:1}) reduces to the classical Smoluchowski (CS) equation as given in (Eq. \ref{eq:smoluchowski})
whose dynamics is governed by the conservation of mass law \cite{ref.smoluchowski}. To obtain a suitable expression for the elapsed time we do a 
simple dimensional analysis in Eq. (\ref{eq:1}) and immediately find that 
the inverse of $\int_0^\infty K(x,y)c(y,t)dy$ is the collision time $\tau(x)$
during which the growth $\alpha x$ takes place \cite{ref.maslov}. 
The mean growth velocity between collisions therefore is
\begin{equation}
\label{eq:2}
 v(x,t) = {{\alpha x}\over{\tau(x)}}=\alpha x\int_0^\infty dyK(x,y)c(y,t).
\end{equation} 
We choose a constant aggregation kernel to make the collision independent of the size of the colliding
particle i.e.,
\begin{equation}
\label{eq:3}
 K(x,y) = 2,
\end{equation}
for convenience.

\begin{figure}

\centering

\subfloat[]
{
\includegraphics[height=7.0 cm, width=7.5 cm, clip=true]
{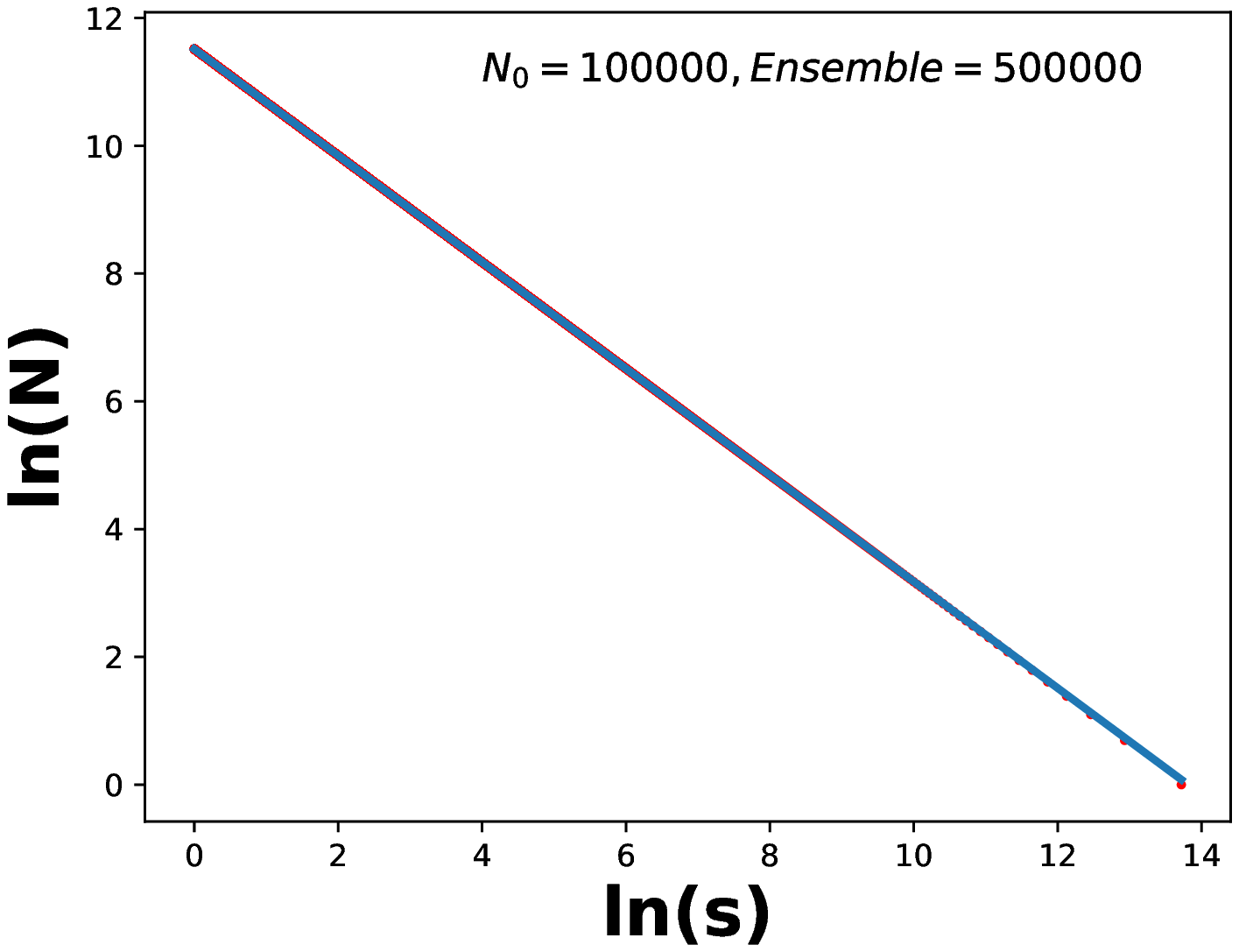}
\label{fig:3a}
}

\subfloat[]
{
\includegraphics[height=7.5 cm, width=8.0 cm, clip=true]
{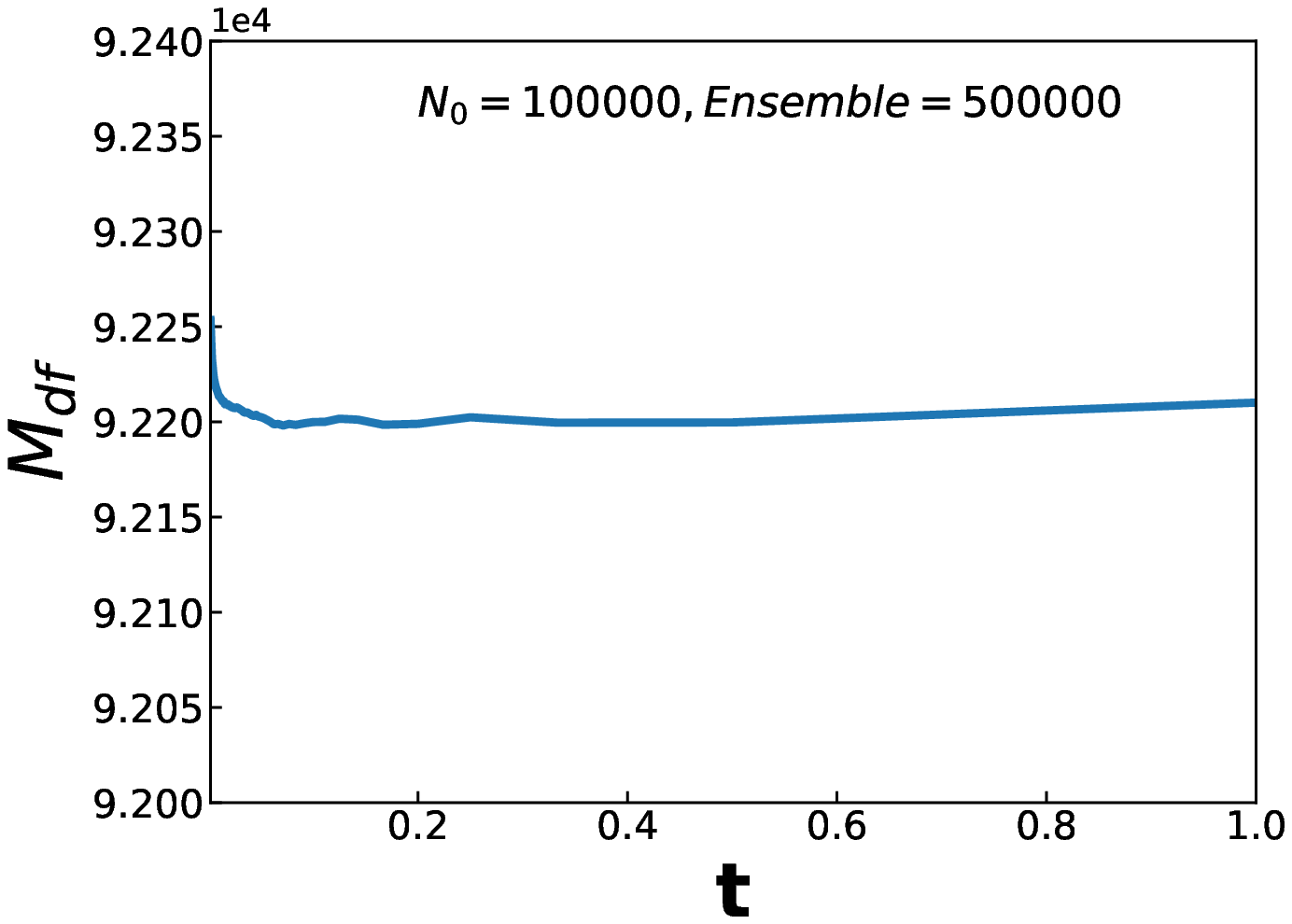}
\label{fig:3b}
}

\caption{(a) We plot number $N$ versus mean particle size $s$ and find excellent straight line
with slope $d_f=1/(1+2\alpha)$ that agrees perfectly with analyitcal solution. (b) The
plot of the $d_f$th moment of $c(x,t)$ versus time $t$ shows it is indeed a conserved quantity. Here, $N_0$ is the initial number of mono-disperse particles and the data 
is averaged over $E=10^4$ number of independent realizations (ensemble size).
} 
\label{fig:3ab}
\end{figure}

\begin{figure}

\centering

\subfloat[]
{
\includegraphics[height=7.0 cm, width=7.5 cm, clip=true]
{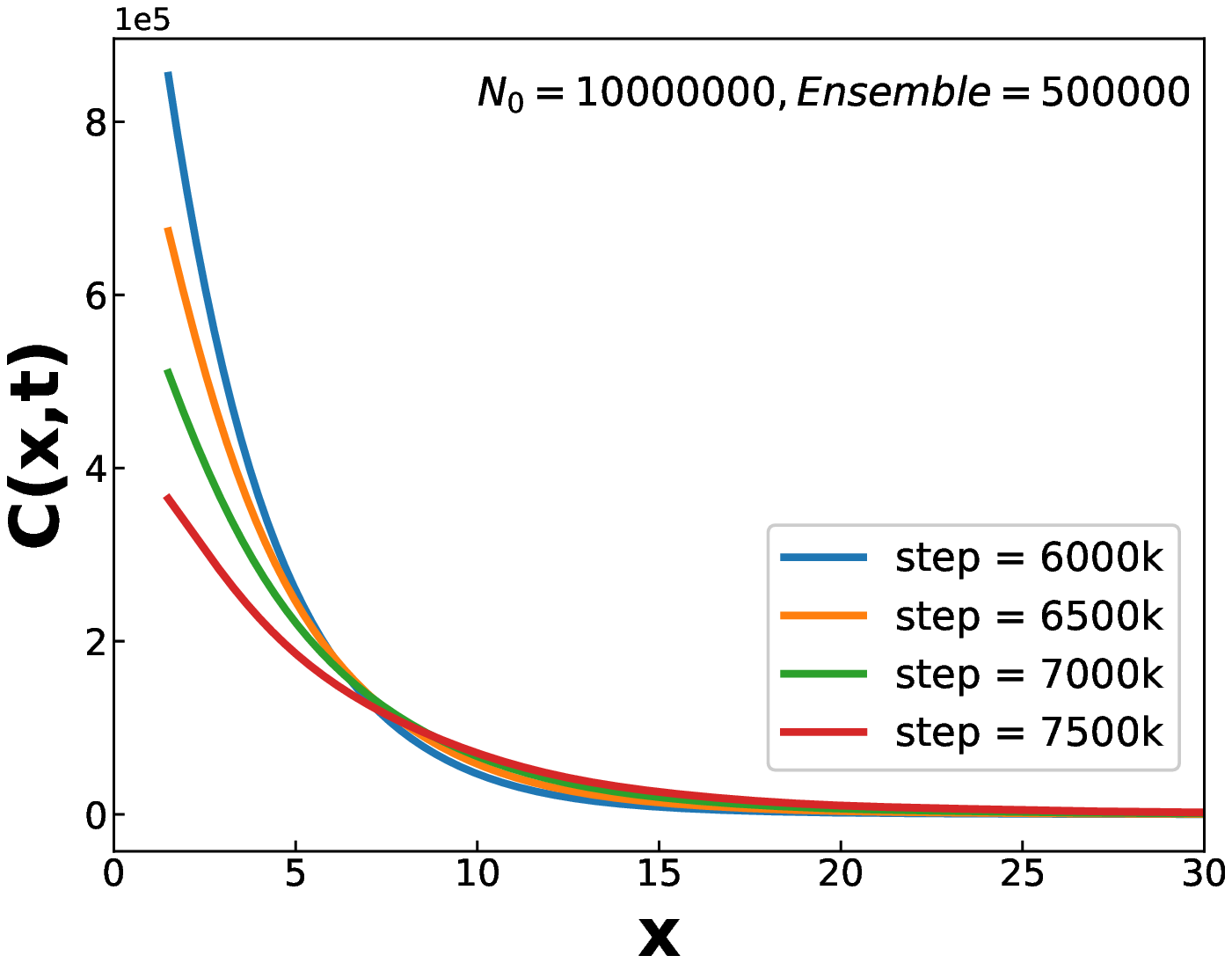}
\label{fig:4a}
}

\subfloat[]
{
\includegraphics[height=7.0 cm, width=7.5 cm, clip=true]
{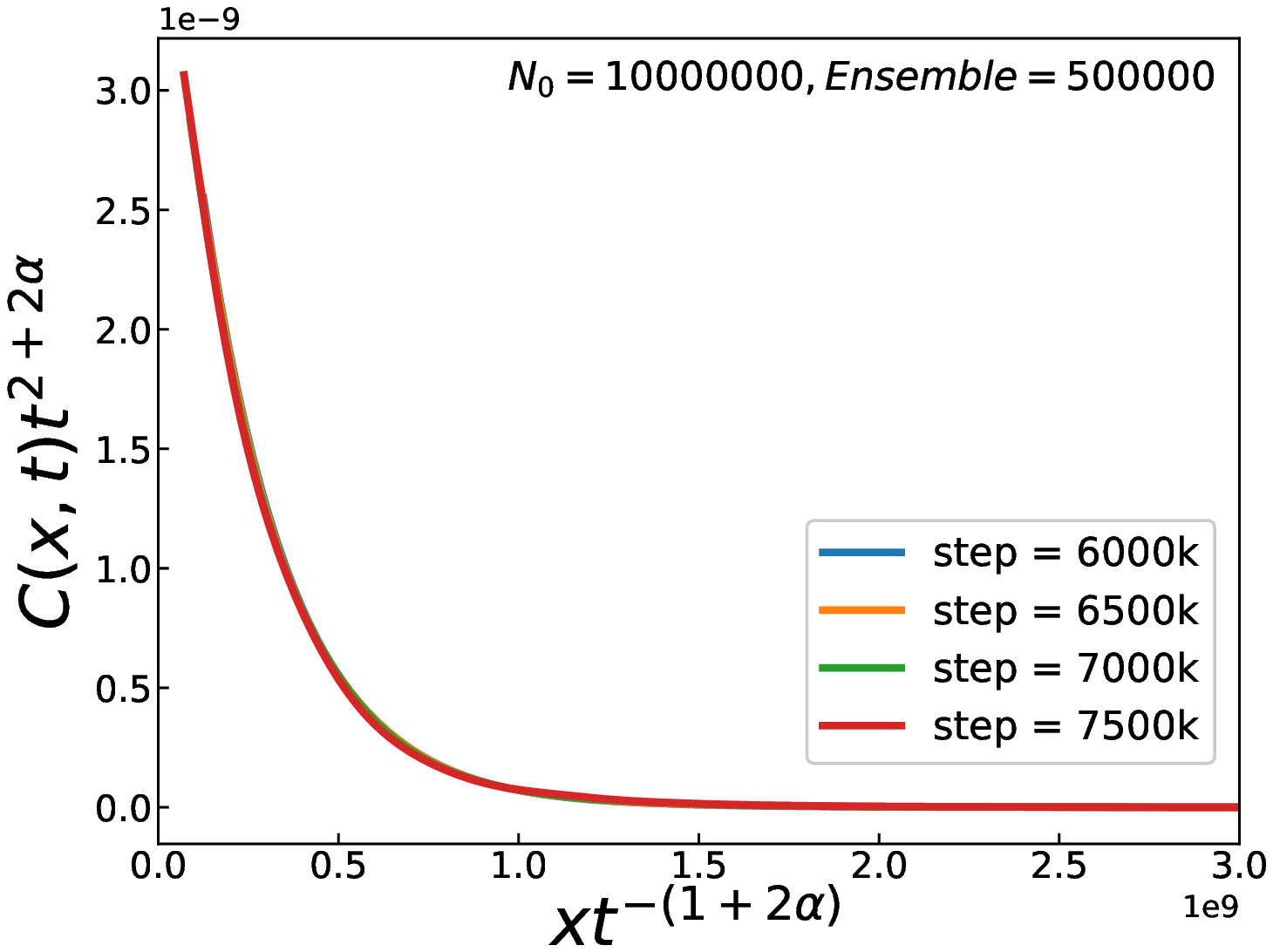}
\label{fig:4b}
}
\caption{
(a) The distribution function $c(t,x)$ is drawn as a function of $x$ for different times.
The same set of data set of (a) are used to plot in them in the self-similar scale (b)
and find excellent data collapse revealing $c(x,t)$ exhibits dynamic scaling.
} 
\label{fig:4ab}
\end{figure}

We can solve Eq. (\ref{eq:1}) for constant kernel with condensation velocity given by Eq. (\ref{eq:2})
analytically and solve it numerically using the following algorithm. 
\begin{itemize}
\item[{\bf(i)}] The process starts with $N$ number of particles of equal sized particles (however
we can choose any distribution since the results are independent initial condition).
\item[{\bf(ii)}] Two particles are picked randomly from the system to mimic random collision via Brownian motion.
\item[{\bf (iii)}] The sizes of the two particles are increased by a fraction $\alpha$ 
of their respective sizes in the logbook to mimic the growth by condensation. 
\item[{\bf (iv)}] Their sizes are combined to form one particle to mimic the aggregation process. 
\item[{\bf (v)}] The logbook is updated by registering the size of the new particle 
in it and at the same time deleting the sizes of its constituents from it.
\item[{\bf (vi)}] The steps (ii)-(v) are repeated {\it ad infinitum} to mimic the time evolution. 
\end{itemize}

Extensive Monte Carlo simulation gives that mean particle size $s(t)$ grows
following a power-law
\begin{equation}
    s(t)\sim t^{(1+2\alpha)}.
\end{equation}
If the size of the total size of the aggregates are measured using mean particle size as an 
yard-stick we find that this number decays with $s$ following a power-law
\begin{equation}
    N(s)\sim s^{-d_f},
\end{equation}
where $d_f$ is the fractal dimension given by
\begin{equation}
\label{eq:30}
d_f={{1}\over{1+2\alpha}},
\end{equation}
as shown in Fig. (\ref{fig:3a}). 
We also find that $d_f$th moment of $c(x,t)$ is a conserved quantity and our numerical
simulation also confirm this in Fig. (\ref{fig:3b}). Note also that the distribution
function $c(x,t)$ obeys dynamic scaling
\begin{equation}
\label{eq:dynamic_s_aggregation}
    c(x,t)\sim t^{-(1+d_f)z}\phi(x/t^z),
\end{equation}
where $z=1+2\alpha$ and $\phi(x)$ is the scaling function \cite{ref.hassan_ba_dc}. 
It implies that the distinct plots as shown in Fig. (\ref{fig:4a}) of $c(x,t)$ versus $x$ 
at different time should collapse into universal curve. To prove
this we then plot $c(x,t)t^{(1+d_f)z}$ versus $x/t^z$ in Fig. (\ref{fig:4b})
and find an excellent data collapse.

The second
model was on the kinetics of aggregation of Brownian particles with stochastic 
self-replication \cite{ref.fractal_self}. It is a very simple variant of the Smoluchowski 
equation in which we 
investigate aggregation of particles accompanied by self-replication of the newly formed
particles with a given probability $p$. That is, we may consider that the system has two 
different kinds of particles: active and passive. As the systems evolves, active particles always
remain active and take part in aggregation while the character of the passive particles are altered irreversibly to an active particle with probability $p$. Once a passive particle turns into an active particle it can take part in further aggregation like other active particles already
present in the system on an equal footing and never turns into a passive particle. The
definition of the model in some sense is reminiscent of the dyadic Cantor set as we just 
have to replace the co-factor ${{1}\over{2}}$ of the gain term 
of Eq. (\ref{eq:smoluchowski}) by ${{1+p}\over{2}}$. 

Like the previous model, the results of this model can be generalized if we express everything in terms of $d_f$ and $z$. The only way, the two results differs are in the value of $d_f$ and $z$ as we find that
$d_f=(1-p)/(1+p)$ and $z=(1+p)/(1-p)$. The results from the two opposing
phenomena reveal that fractal emerges only if the total mass of the system grows with time 
following a power-law in aggregation and in fragmentation process fractal emerges only if the total mass decreases with time. 
The opposite reasons 
for the emergence of fractal reflects the inherent opposing nature of the two phenomena. 
In all these kinetic systems, the emergence of fractal is always accompanied by the 
conservation of the $d_f$th moment of $c(x,t)$  \cite{ref.fractal_conserved}. 

\section{Kinetics of random sequential adsorption process}

Random sequential adsorption (RSA) is yet another field of research where Robert Ziff worked quite 
extensively \cite{ref.rsa_ziff_1,ref.rsa_ziff_2,ref.rsa_ziff_3}. One
of the results that I really liked in RSA is the jamming limit for RSA of monodisperse particle, 
which is also known as car-parking problem, in one dimension as it can be solved exactly
\cite{ref.car_parking}. It has been 
found that the fraction of the total space being occupied by the depositing particles 
is $0.747 597 920 3 . . $,
provided sizes of the depositing particles are negligibly smaller than the size of the substrate
\cite{ref.renyi}. 
However, most of his work has been on two dimensional substrate which is physically more relevant than
its one dimensional case. In two dimensions, works are mostly done by numerical simulation. 
I was interested more on analytical solution and hence concentrated only on one dimension. I solved RSA
model for a mixture of two different sized particles, a mixture of particles that follows
a power-law and a mixture of point like and fixed sized particles 
\cite{ref.hassan, ref.hassan_rsa1,ref.hassan_rsa2}.

\section{Thermodynamic formalism of percolation}

The field of research where Robert Ziff has contributed the most is the theory of percolation
\cite{ref.per_ziff_1,ref.per_ziff_2,ref.per_ziff_3,ref.per_ziff_4}. Percolation
model is perhaps one of the most elegant concept in statistical physics which can be used in many 
different situations of science, arts and social science. Despite being first conceived
in 1941 by Flory, its mathematical formulation was first given in 1957 by Broadbent and Hammersley
\cite{ref.Broadbent}. 
Since then it has been one of the most studied models. As far as the definition of percolation model is concerned everyone would agree that it is 
one of the simplest models in statistical physics \cite{ref.Stauffer}. On the other hand, 
it is also one of the 
hardest models in statistical physics since we only have a few exact solution in dimensions 
$1<d<\infty$. The idea of percolation was first
conceived in the early 1940’s by chemist Paul Flory in his study on gelation in polymers, although he
did not use the word "percolation" \cite{ref.Flory}. The very word "percolation" was first used
and its mathematical formulation was first proposed by engineer Simon Broadbent and mathematician John Hammersley in 1957 \cite{ref.Broadbent}. In their seminal paper, the authors clearly stated that their work has the
potential to encourage others to investigate this terrain, which has both pure mathematical
fascinations and many practical applications. Indeed, the following years, namely the entire 60s decade,
the works of a group of celebrity researchers like 
Cyril Domb, Michael Fisher, John Essam, and M.F. Skyes, Rushbrooke, Stanley, Coniglio, Halperin, Herrmann, Stauffer, 
Aharony, Havlin, Duplantier, Cardy, Grassberger and Ziff
popularized the percolation problem among both physics and mathematics communities by
establishing percolation as a critical phenomena \cite{ref.domb_1, ref.domb_2, ref.domb_3, ref.essam_1,
ref.essam_2}. 

The simplest way to define percolation is to first choose a lattice or network as a skeleton
and assume that initially all the disconnected sites are present so that each site
is a cluster of its own size. As we occupy bonds, we find that
for small occupation probability $p$ there are only small isolated clusters that do not span 
across the entire lattice and if $p$ is close to one there will definitely a cluster which would
span the lattice. Thus, there exists a critical or threshold value
so that in the thermodynamic limit 
at $p>p_c$ the system will at least have one cluster that spans across the linear size of the lattice
\cite{ref.Stauffer, ref.saberi}.
Finding percolation thresholds $p_c$ both exactly and by numerical simulation has been an enduring
subject of research in percolation. Robert Ziff found critical points for many different lattices
and  made significant contributions to exact solutions both on the lattice level, in the continuum 
limit and on the networks too \cite{ref.per_ziff_1,ref.per_critical_1,ref.per_critical_2}. 
Besides, finding an efficient algorithm for percolation is always considered significant development
as that would mean we can simulate for larger system size and get results for higher ensemble size.
To this end, the first and most classic algorithm is the Hoshen and Kopelman algoritthm and later 
the Leath algortithm \cite{ref.hoshen, ref.leath}. However, the last and currently most efficient algorithm is the Newman and Ziff algorithm \cite{ref.per_ziff_0,ref.per_ziff_algorithm}. It 
measures an observable quantity in a percolation system for the
full spectrum of occupation probability $p$ in an amount of time that scales
linearly with the system size. In that article, Newman and Ziff also showed how to get an observable
as function of $p$ from the data as a function of the number of occupied bonds $n$ which
gives the results in the canonical ensemble. The data obtained in the canonical ensemble makes
the ensemble average computationally much cheaper and smoother.

One of the major breakthrough was made by Kasteleyn and Fortuin in 1969 who showed that 
q-state Potts model 
in the limit $q\rightarrow 1$ corresponds to percolation model \cite{ref.Kasteleyn}. It paved the way to identify the 
equivalent counterpart of order parameter and susceptibility in percolation which was crucial
to prove that percolation is indeed a model for second order phase transition and to find the
corresponding critical exponents. The best known example of the second order phase transition
is the paramagneitc to ferromagnetic transition or vice versa and the
simplest model that can capture its various aspects successfully is the Ising model. 
Despite its simplicity, its exact solution in two or in higher dimensions remained an open problem
for many years until Onsager solved it exactly. However, a more physical understanding 
of the phenomena was made by Wilson, Fisher and Kadanoff who showed that the system at and
near the critical point is scale invariant \cite{ref.kadanoff,ref.wilson, ref.fisher}. It means that if we blow up the picture near
the critical point by some factor $b>1$ then it would look the same, at least
in the statistical sense. This simple idea played a crucial 
role for the renormalisation group which led to a deeper understanding of the critical
phenomena including percolation. Soon, Alexander Polyakov established a connection 
between the idea of scale invariance and the conformal invariance \cite{ref.polyakov}. Polyakov argued 
that the picture of the system should also remain statistically similar if the 
factor $b$ is allowed to vary smoothly as a function of the called conformal 
mappings. Stanislav Smirnov was awarded Fields Medal in 2010 for his proof of conformal invariance at
criticality \cite{ref.smirnov}. 

I was so much inspired by the work of Robert Ziff on percolation that I always wanted to do 
something with it. Teaching is the best way to learn a subject. To that end, in 2005 I designed a course
titled "Non-equilibrium Statistical Mechanics" and included percolation in the chapter of
phase transition and critical phenomena. That was the time I really began to understand the subject.
That is the time when I also realized that despite the 60-year-long active research on percolation
we still did not know how to define entropy. This is one of the most crucial quantity for defining
the order of transition since without it, we can never know specific heat and whether the transition involves any latent heat or not. Besides, I also realized that the critical exponents of percolation 
must also obey the Rushbrooke inequality. In other words, for every quantity that we have in
thermal phase transition we must have an exact equivalent counterpart in percolation. 

The first question I wanted to explore is: What if I replace a regular planar lattice
with a  weighted planar stochastic lattice 
(WPSL) as a skeleton. We proposed this lattice in 2010 and shown that it is multifractal and, unlike 
square or regular lattice, its coordination number is not fixed rather exhibits a power-law. I engaged
one of my MS thesis students to work on this problem in 2014 and in 2015 we had our 
first article published in the Physical Review E as a rapid communication \cite{ref.hassan_mijan_1,ref.hassan_mijan_2}. One of the interesting findings of 
phase transition and critical phenomena in general
is the universality class. It is one of the central predictions of renormalisation group theory that the 
critical behaviours of many statistical mechanics models
on Euclidean lattices depend only on the dimension and not on the specific choice of lattice. 
As far as regular Euclidean lattice is concerned, it has
been found that the critical exponents of percolation too depends only on the dimension of
the lattice and are independent of the details of the lattice. Our work on WPSL suggests
 that it is, however, not true if the planar lattice is scale-free and multifractal. 
This is perhaps the only
known exception where despite its dimension is the same as that of the square or triangular lattice
yet it belong to different universality class.

Percolation is well known as a model for second order phase transition since the later part of the 
60s decade. Second order phase transitions occur when a new state of reduced symmetry develops 
continuously from the higher symmetric disordered (high temperature) phase. In fact, 
two quantities are crucial to determine the order of transition: order parameter and entropy. Order parameter quantifies the extent of order and entropy measures the degree of disorder. A full descriptions of phase transition is only possible
if we know the behavior of both the quantities. However, in the last 60 years enough attention 
has not been paid to know how to measure entropy for percolation. To the best of our knowledge, 
the first attempt to obtain entropy in percolation was made by Tsang and Tsang in 1999 and a slightly different definition was used by
Vieira {\it et. al} in 2015 \cite{ref.tsang,ref.tsang_1,ref.Vieira}. They both found that entropy is maximum at the
critical point and zero at the initial state. We all know that order parameter too is zero 
at the initial state and remain zero till the critical point. However, due to finite-size
effect order parameter can be non-zero but small near the critical point. However, the behavior
of order parameter for increasingly larger system size clearly shows
the sign of becoming zero in the thermodynamic limit  up to the critical point. 
If initially entropy and order parameter are both equal to zero it means that the system
is at the same time in a perfectly ordered and disordered state. This cannot be right and
hence something must be wrong in their choice of probability to define entropy. 
That was the time when I started my 
journey to find a proper way of obtaining entropy for percolation. 

In 2017, we were successful 
as we used cluster picking probability $\mu_i=s_i/N$ where $s_i$ is the size of the $i$th cluster
and $N$ is the system size \cite{ref.hassan_didar,ref.hassan_shahnoor}. It describes the probability 
that a site, picked at random, belongs to the
$i$th cluster. Using this probability in the definition of Shannon entropy
\begin{equation}
\label{eq:shannon}
    H=-K\sum_i\mu_i\log\mu_i,
\end{equation}
where the constant $K$ just amounts to a choice of a unit of measure of entropy and hence
we choose $K=1$ for convenience  \cite{ref.shannon}. It gives the desired entropy which is 
consistent with the nature of order parameter. Substituting $\mu_i=s_i/N$ in Eq. (\ref{eq:shannon}) we get
\begin{eqnarray}
 H &= & -{{1}\over{N}}\sum_i s_i\log s_i+{{1}\over{N}}\sum s_i\log N \nonumber  \\ 
&=& {{1}\over{N}}\log {{N!}\over{s_1!s_2!....s_k!}} \nonumber = {{1}\over{N}}\log\Omega,
\end{eqnarray} 
where $\Omega$ is the number of distinct ways $N$ number of sites can be arranged
into $k$ number of clusters of sizes $s_1,s_2,...,s_k$. Thus, the Boltzmann entropy 
for that specific state is $S= k_B \ln \Omega$. If we choose the Boltzmann constant $k_B=1$,
for convenience, since
it merely amounts to the choice of the measure of entropy then we find that Shannon entropy 
actually is the entropy per site 
In other words, the Shannon entropy is the specific entropy and 
the total entropy $S$ is equal to $S=NH$ which makes it an extensive property as it should be.

\begin{figure}

\centering

\subfloat[]
{
\includegraphics[height=7.0 cm, width=8.0 cm, clip=true]
{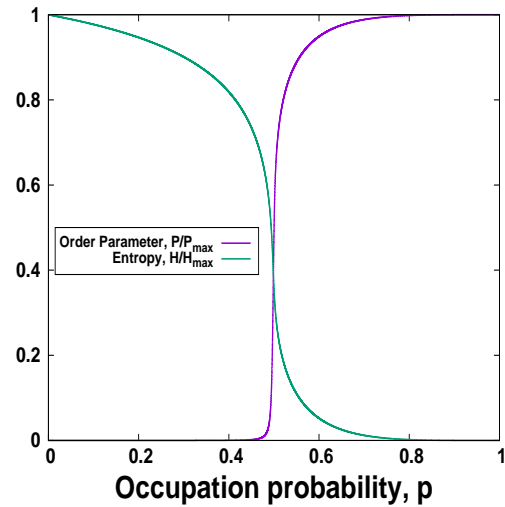}
\label{fig:5a}
}

\subfloat[]
{
\includegraphics[height=7.0 cm, width=8.0 cm, clip=true]
{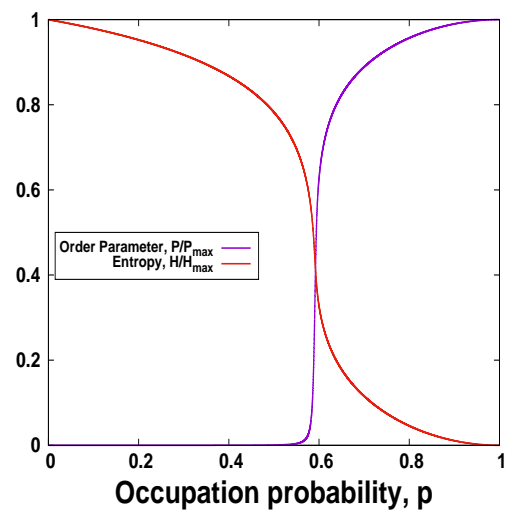}
\label{fig:5b}
}

\caption{Plots of scaled order parameter $P(p)/P(1)$ and scaled entropy $H(p)/H(0)$ 
as a function of occupation probability $p$, fraction of bond (site) being occupied, 
for lattice size of linear size $L$ are shown in the same
plot for  (a) bond (b) for site percolation. In (a) both cases the two curves 
meet almost at the critical occupation probability $p_c$.
} 
\label{fig:5ab}
\end{figure}

We start the percolation process such that initially every site is isolated
and hence there are $N=L^2$ number of equal sized cluster of size one in the square lattice
of linear size $L$. Initially, $\mu_i=1/N \ \forall \ i$ and hence
substituting it in Eq. (\ref{eq:shannon}) yields $H(0)=\log(N)$ which is
the maximum possible entropy. This situation is analogous to ideal gas which corresponds to the most disordered state since the entropy is maximum. In percolation, the maximum entropy at $p=0$ is consistent
with the behavior of order parameter $P$ since we have $P=0$ at $p\leq p_c$. At other extreme when
all the sites are connected we have $\mu_i=1$ and thus entropy is equal to zero. This state can be 
regarded as analogous to crystal state where there is just one perfectly ordered microstate.
This state of $H=0$ is again consistent with the behavior of order parameter as we find
that the order parameter $P$ is maximally high. Thus, it suggests that the system is almost 
in the perfectly ordered state. The second order or continuous phase transition
is also known to be accompanied by an order-disorder transition although in percolation this aspects 
of phase transition have never been looked at. Plotting entropy and order parameter 
in the same graph can provide better insights into this. To this end, we plot
 both $H$ and $P$, though we re-scaled both entropy and order parameter by $H(0)$ and $P(1)$ 
 respectively, in the same graph for random bond and site percolation as shown in 
 Figs. (\ref{fig:5a}) on square lattice.
We see a sharp and sudden change of both quantities near $p_c$ and they both meet almost
at $p_c$. This is a sign of order disorder transition as we see
that at $p<p_c$ order parameter $P$ is minimally low where entropy is maximally high and vice versa
at $p>p_c$. Thus, the  $H-P$ plot suggests that the transition is accompanied by an order-disorder transition \cite{ref.hassan_didar,ref.hassan_shahnoor}.

While measuring entropy for site percolation on square lattice we have found that it suffers a problem. 
In the traditional definition of site percolation we occupy a site and we measure the cluster size in
terms of the number of contiguous occupied sites. On the other hand, in bond percolation we occupy a site
and measure cluster in terms of the number of contiguous sites connected by the occupied bonds. That is,
there are two entities, namely bonds and sites, such that when bonds are occupied
we must connect sites and vice versa. This has not been the case for the traditional definition of 
site percolation. We therefore recently redefined site percolation \cite{ref.hassan_shahnoor}. We assume that initially
all the isolated bonds are already present in the system. The occupation of sites then connects
the bonds and cluster sizes are measured by the number of contiguous bonds connected by the occupied 
sites. Interestingly, the new definition of site percolation does not change the values of the
critical point and the critical exponents except now the behavior of entropy is consistent with the
order parameter. We now show the plots of order parameter and
entropy in Fig.  (\ref{fig:5b}) according to the new definition of site percolation. The 
plots of $H$ and $P$ meet near the critical point like for bond percolation suggesting that site 
percolation too is accompanied by order-disorder transition. 

One of the reasons why percolation is so elegant and has been well studied for more than 60 years is that
it has many features in common with its thermal counterpart. One of the features is definitely the universality. We know that the critical exponents in thermal phase transition depend only on the
dimension of the lattice or system, the spin dimensions and the extent up to which spin can interact. Interestingly,
their values are independent of the detailed nature of the lattice structure and the strength of
interaction. The critical exponents in percolation too are well-known to depend 
only on the dimensions of the lattice or system and independent of the types of percolation, namely
whether the percolation is bond and site type, as well as of the detailed structure of the lattice. However, in 2015 we find
that this is no longer the case if the lattice is multifractal and scale-free as we find that
its coordination
number distribution obeys inverse power-law. That is, we performed percolation on WPSL which is a
planar lattice and hence it was expected to belong to the same university class as that of square
lattice \cite{ref.hassan_mijan_1}. Instead, we, find new set of critical exponents that clearly suggests that scale-free and multifractal nature have an impact in determining the
universality class.

Besides Euclidean regular and scale-free lattice as a skeleton for percolation, the use of network 
(random, scale-free and small-world networks) as a skeleton is gaining increasingly more
popularity. Networks are not embedded in the Euclidean space rather in the abstract space. 
Networks can mimic many real-life systems such as transportation network (like the world-wide airline network) or a communication network (like world wide web network). Besides, 
viruses are typically spread on a social or through computer network and hence whether 
such spreading will cause epidemic or pandemic or will just die out will depend on the nature of its
network architecture. The history of percolation on random graph goes back to the work of
Erd\"{o}s and R\'{e}nyi in 1959 \cite{ref.erdos}. The process starts with $N$ isolated nodes
and we connect them by adding links one at each step after picking them 
randomly. It is intuitively clear that as the number of added links $n$ is increased, 
contiguous occupied nodes keep forming clusters and their average size 
continue to grow. Even in such simple process we observe that suddenly a giant cluster, whose
size is $O(N)$, emerges across a critical point $t=1/2$ where the relative link density $t=n/N$. Note that in the stochastic processes like fragmentation and aggregation $t$
is regarded as time. It is also important to note that in the context of percolation on network $t$ plays the same role as occupation probability $p$ of percolation on lattice.
Erd\"{o}s and R\'{e}nyi demonstrated that below $t_c$ 
even the largest cluster $s_{{\rm max}}$ is of miniscule size $O(\log N)$  
\cite{ref.erdos, ref.bollobas}. 
The transition from such miniscule to a giant cluster is called percolation which is analogous to the 
transition from non-spanning to spanning cluster in a spatially embedded lattice.

\begin{figure}

\centering

\subfloat[]
{
\includegraphics[height=7.5 cm, width=7.0 cm, clip=true, angle=-90]
{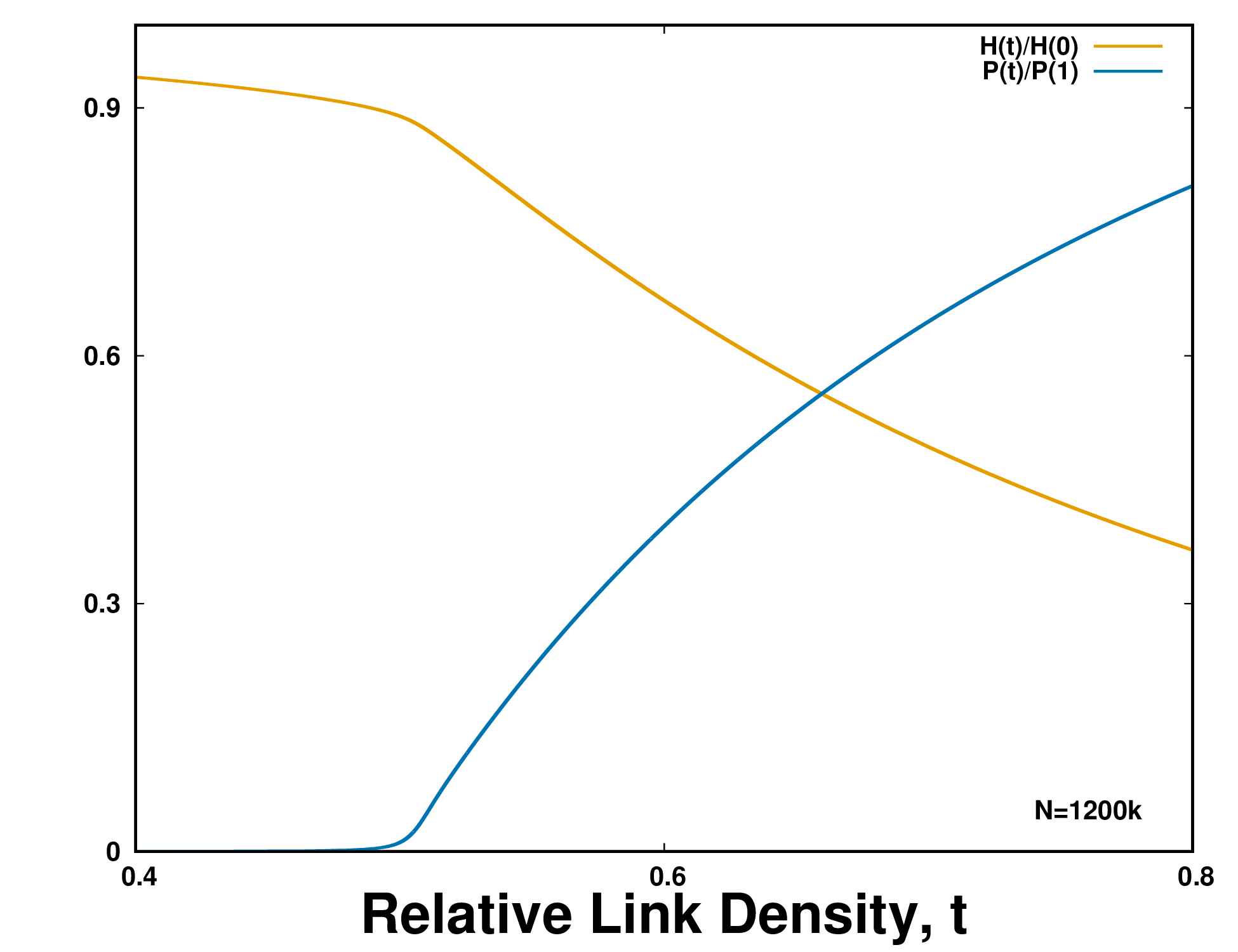}
\label{fig:6a}
}

\subfloat[]
{
\includegraphics[height=7.5 cm, width=7.0 cm, clip=true, angle=-90]
{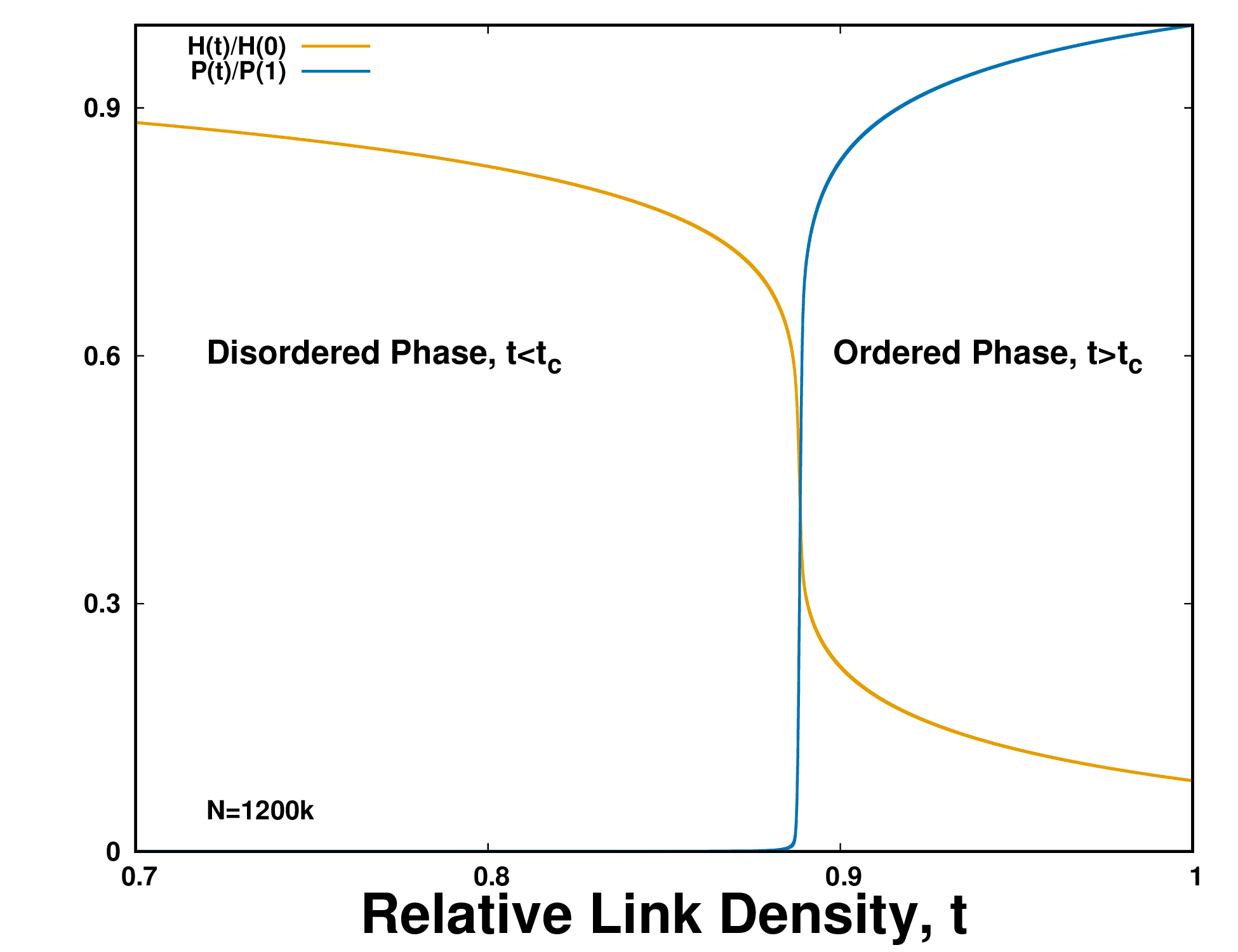}
\label{fig:6b}
}

\caption{Plots of scaled order parameter $P(t)/P(1)$ and scaled entropy $H(t)/H(0)$ 
as a function of $t$ for fixed network size $N$ are shown in the same
plot for  (a) ER (b) for EP. 
} 
\label{fig:6ab}
\end{figure}

It is true that percolation in random or Erd\"{o}s and R\'{e}nyi graph describes a phase transition.
However, the order transition and the nature of symmetry breaking have never been clearly stated. 
Our plots of entropy and
order parameter for percolation on ER network suggests that above $t_c$ they both are sufficiently 
high as seen in Fig. (\ref{fig:6a}). It means above $t_c$ the system is moderately ordered and moderately disordered at the same time. It also means that ER transition is clearly not an
order disorder transition. The situation changes, however, if we pick two links randomly instead of one at
each step but occupy only the one that forms a smaller cluster than the other link, which
we then discard. As the growth of the larger cluster is systematically discouraged, the transition
is delayed which is expected but when it happens it undergoes a transition in an explosive fashion
and hence it was termed as an "explosive percolation" by Achlioptas {\it et. al} in 2009
\cite{ref.Achlioptas}. Initially, it was mistaken as discontinuous 
or first order transition but finally in 2011 it was shown to describe second order transition.
This time the order parameter $P$ 
and entropy $H$ undergo a much sharper transition than the ER transition.
It can be easily seen in Fig. (\ref{fig:6b})that where $P$ is minimally low and $H$ is maximally high and vice versa suggesting it is truly an order-disorder transition.

The question is: Why do we get sharp order-disorder transition when we encourage smaller cluster to grow
faster than the larger cluster? Friedman and Landsberg argued that encouraging the smaller cluster to 
grow faster helps the system develops {\it powder keg} and eventually the mitigation of this effect 
results in the explosive percolation which is called {\it powder keg} \cite{ref.Friedman}. 
However, the exact physics 
behind this {\it powder keg} effect has not been explained yet. We know the expression for free 
energy $F=E-TS$ from thermodynamics. In a given situation the minimum of $F$ always 
corresponds to the stable state. Due to the competing role of $E$ and $S$, the minimum of 
$F$ can be achieved in two ways: In the disordered phase, the minimum of $F$ is 
achieved by maximizing $S$ and in the low temperature phase the minimum of $F$ 
is achieved by minimizing $E$. In percolation, initially at $t=0$ the system 
is at its maximum entropy or at its utmost disordered phase. Thus, when there are options available the system will choose the 
one that minimizes $F$. In explosive percolation, adding the link that forms smaller cluster helps 
the system continue to stay at a higher entropy state.
Note that increasing the number of available options makes the system 
stays longer at a higher entropy state and hence the onset of transition
is delayed further thus making the transition at a higher $t_c$ value. However, regardless of the 
number of options we choose, the critical point $t_c$ never exceeds one. It suggests that we can consider $1-t$ as the equivalent counterpart of temperature.

\section{Discussion and conclusions}

In this article, I have tried to outline the research field of Robert Ziff alongside
my contribution to the fields as motivated by his works. I started with the kinetics of 
fragmentation where
Ziff obtained many exact solutions to the fragmentation equation. The most significant finding
of him in fragmentation is the shattering transition where the smaller particles breaks at an
increasingly rapid rate that results in the loss of mass to the "zero size" phase. This is in some
sense reminiscence of Bose-Einstein condensation. I then discussed how we can use the fragmentation
equation in one dimension to describe fractal and in higher dimensions multifractal. 
A system that evolves in time in such a way that all snapshots are similar in the same sense
as two or more triangles are similar is said to obey dynamic scaling. Self-similarity in continuous
time axis is also a kind of continuous symmetry. It suggests that
there must be a conserved quantity as a controlling agent behind preserving this 
self-similarity along the time axis. When the total mass is a conserved quantity
 the system is not a fractal. When the mass of the system is not a conserved quantity
 the system is a fractal provided there still exists a conserved quantity required by the
 self-similar symmetry which is reminiscence of Noether's theorem. The conserved quantity
 in the $d_f$th moment of the
 fragments size distribution function $c(x,t)$. 
 
 I then discussed about the opposite of fragmentation phenomena namely the kinetics of aggregation in which Ziff has made significant contribution too. Apart from finding the scaling solution to the Smoluchowski equation, the most significant finding
 is that under certain conditions it can describe gelation transition which is just opposite
 to shattering transition. I have shown that the Smoluchowski equation can describe fractal too
 but only if the total mass of the system grows with time following a power-law. The resulting
 system also exhibits dynamic scaling - a litmus test of continuous self-similarity along time axis. Like
 the emergence of fractal in fragmentation, here too we find that the $d_f$th moment of $c(x,t)$ is a conserved quantity which essentially controls the self-similar symmetry along time axis. Thus, the emergence of
 fractal in both aggregation and fragmentation shares some common features. Namely, (i)
 they both exhibit dynamic scaling, (ii) total mass in fragmentation must decrease and increase
 in aggregation but in both the cases they do it following a power-law.

 Finally, I discuss the various new aspects of percolation theory where Robert Ziff's contribution
 is truly outstanding. Percolation is the simplest model in statistical physics that describes
 second order phase transition. Real thermal physical system that exhibits second order phase transition is also accompanied by order-disorder transition. It also means symmetry is broken as the system undergo transition from disordered to order which forms the heart of
Landau theory of phase transition. For more than 60 years we only knew how to measure the extent of
order as we could measure order parameter for percolation. To know whether the transition also 
breaks symmetry or not we must know how to measure entropy for percolation. Recently, I have defined entropy for percolation which paved the way
 to define specific heat too. Furthermore, I have recently redefined susceptibility since the critical 
 exponent $\gamma$ obtained from mean cluster size was too high to obey Rushbrooke inequality with
 positive value of $\alpha$ of the specific heat. The critical exponents obtained from the redefined quantities obey the Rushbrooke inequality for random and explosive percolation on network and lattice
 both. Besides, I have discussed about the role of entropy in checking whether percolation is
 also accompanied by order-disorder transition. To that end, it has been found that percolation on 
 ER network does not break symmetry or is not accompanied by order-disorder transition unlike explosive 
 percolation. 
 
 We all know that gravitational forces not only act on things that are close together but they
also act on things that are at great distances apart. The Earth and Moon are 384,400 km apart and yet
gravity creates an immense force which is often referred to as action at a distance. Like
wise researchers can create a generation of new researchers through their published works which 
can also be referred as action at a distance. We do PhD under a direct supervisor but we often
fail to recognise that we are also being supervised by many action at a distance supervisors.
Through this Robert Ziff and many others like him have been influencing new generation of researchers.
In this article I tried to emphasize this. In conclusion, I have given a comprehensive account of fragmentation, aggregation and percolation processes
and tried to outline their key features. I dedicate this paper to Robert Ziff on the occasion
of his 70th birthday whose scientific leadership and influential works have been
continuously guiding, inspiring, and challenging my generation and will continue to do 
so to the generation to come.

\end{document}